\documentstyle[epsfig,longtable]{aipproc}

\begin{document}

\title{Making Damped Lyman-$\alpha$ Systems in Semi-Analytic Models}

\author{Ariyeh H. Maller$^1$, Rachel S. Somerville$^{2}$,
Jason X. Prochaska$^{3}$ and Joel R. Primack$^{1}$}
\address{$^1$Physics Department, University of California, 
Santa Cruz, CA 95064, USA\\
$^{2}$Racah Institute of Physics, The Hebrew University, 
Jerusalem, 91904, Israel\\
$^3$Observatories of the Carnegie Institution of Washington, Pasadena CA 91101}

\maketitle

\begin{abstract}
The velocity profiles of weak metal absorption lines can be used to
observationally probe the kinematic state of gas in damped Lyman-$\alpha$
systems. Prochaska and Wolfe \cite{pw96}
have argued that the flat distribution
of velocity widths ($\Delta v$) combined with the asymmetric line profiles
indicate that the DLAS are disks with large rotation velocities ($\sim$200
km/s). An alternative explanation has been proposed by Haehnelt, Steinmetz, and
Rauch (HSR)\cite{hsr98}, in which the observed large velocity widths and 
asymmetric
profiles can be produced by lines of sight passing through two or more clumps
each having relatively small internal velocity dispersions. We investigate the
plausibility of this scenario in the context of semi-analytic models based on
hierarchical merging trees and including simple treatments of gas dynamics,
star formation, supernova feedback, and chemical evolution. We find that all
the observed properties of the metal-line systems including
the distribution of $\Delta v$ and the asymmetric profiles, can be reproduced
by lines of sight passing through sub-clumps that are bound within larger
virialized dark matter halos. In order to produce enough multiple hits, we find
that the cold gas must be considerably more extended than the optical radius of
the proto-galaxies, perhaps even beyond the tidal radius of the sub-halo. This
could occur due to tidal stripping or supernova-driven outflows.
\end{abstract}

\section*{Introduction}
Damped Lyman alpha systems (DLAS), by probing the gas content of the 
universe at high redshift, are powerful observational probes 
to study galaxy formation and evolution. Until recently the observations
of DLAS included the differential density distribution $f(N)$, its
evolution with redshift, 
and the metalicity of the absorbers \cite{storie96}.  
Many authors have shown that 
hierarchical cosmologies with models of galaxy formation can match this data.
Most notable is the work of Kauffmann \cite{kauf96} 
which uses semi-analytic models
(SAMs) to match $f(N)$ and its evolution with redshift, and at the same time 
reproduces 
many features of galaxies observed locally in emission. A general prediction 
of Kauffmann's models is that the galaxies producing DLAS will typically be
smaller then galaxies today. 

Prochaska and Wolfe \cite{pw97,pw98}, hereafter PW, introduced new data by
studying the kinematic properties of DLAS as deduced from high
resolution spectroscopy of their weak metal lines.  They found that DLAS
seem to have a wide range of velocity widths, and are asymmetric in
velocity space.  The only model they tried that could
explain this had thick disks with rotational velocities $\sim 220$ kms$^{-1}$,
which is incompatible with what is expected in models of hierarchal
structure formation at the typical redshifts $z\sim3$ of these
observed DLAS.  PW excluded the hierarchical single-disk model of
Kauffmann.  

Recently HSR showed using hydrodynamic simulations and the
Press-Schechter approximation that this seeming incompatibility can be
reconciled if the DLAS come from merging proto-galactic clumps.
Then the DLAS are composed of a few objects, 
and the large velocity widths come from the motions of
these objects within a larger dark matter halo.  
McDonald and Miralda-Escud\'{e} \cite{mm98} 
have tested this in an analytic model.  We set out to
investigate if such a model is feasible in the context of SAMs, which
have allowed us to include star formation and feedback, and ultimately
to include also a wider range of CDM-type cosmological models.

\section*{The Semi-Analytic Models}
SAMs and hydrodynamical simulations have their individual strengths and 
weaknesses.  Hydrodynamical simulations include 
gas dynamics and gravitation, but this makes them computationally 
expensive and thus limited in their ability to explore parameter space
or even get adequate statistics from the small volume simulated at high
resolution. 
Also because of this it can be difficult to understand in simple terms 
what is going on in the simulation. Thus while HSR showed
that their model is consistent with the PW kinematic data, 
they only simulated
one cosmology and they did not include feedback or metal production.

In contrast, SAMs attempt to capture the most essential aspects of galaxy
formation, in a simplified manner. Thus instead of following the gravitational
motion of particles a merger history is assigned statistically to a halo. Gas
cooling, star formation, and feedback are treated by simple equations.  The
fact that SAMs are capable of matching many observations suggests that these
simpler treatments are successful in reproducing the essential features of
galaxy formation.  We
use the SAMs of Somerville and Primack \cite{sp98} 
and Somerville, Primack and Faber (SPF)\cite{spf98}, 
which are similar to the work of Kauffmann \cite{kauf96}.

In particular,
we use the fiducial model of SPF, which reproduces the observed number
densities of the Lyman break galaxies and $\Omega_{\rm gas}(z)$ and
metalicities $Z(z)$ measured from DLAS.  In previous work such as PW's
analysis of the results of Kauffmann, it was assumed that each observed
velocity width arose from the internal rotation velocity of a single disk. The
main new feature here is that we explicitly investigate the implications of the
substructure in the matter halos for the metal-line kinematics, and different
ways of distributing gas within the proto-galaxies.  If we are to reproduce the
kinematics of DLAS then, the large velocity widths must come from a combination
of the rotational motion of the gas and the relative motion between the objects
in one halo.

\section*{Making the DLAS}

We analyze SCDM ($\Omega_{\rm matter}=1$, $\sigma_8=0.67$) 
with the star formation and feedback described in SFP's fiducial model 
at a redshift of 3.2. This gives us the number of satellites in a given
halo, their distances from the central object, and the amount of cold gas
in each object.  We position the satellites in the halo randomly in
angle along circular orbits.  
The only quantity not specified by the SAMs is how the cold
gas is distributed in each galaxy.  We construct five models for the radial 
distribution of the gas, to explore how the radial profiles affect the observed
kinematics (Table \ref{tableAri}).
We distribute the gas with an exponential of isothermal  radial profile N(R),
and different normalizations.
We compare these models to the data by passing random lines of sight 
through a given halo, with each halo
weighted by the probability of encountering that halo as determined by the
Press-Schechter formalism.  Those lines of sight that intersect cold gas in
excess of 2 $\times 10^{20}$cm$^{-2}$ are labeled as DLAS.  Then the kinematics
of these systems are investigated by assuming the gas disks have metalicities 
given by the SAMs and a scale height of one tenth the stellar disks scale
length.  Spectra are simulated and then analyzed by the same four statistics 
as the data was in PW, the velocity width $\Delta V$, the mean to median test 
$f_{mm}$,
the edge leading test $f_{edg}$, and the two peak test $f_{tpk}$.
We also check to see that
we are getting a reasonable distribution of column densities $f(N)$ 
(Figure 1a). All models but the first and third 
fit the DLAS $f(N)$ reasonably well.     

\section*{Conclusions}

Of the models we explore only the last model comes close to matching all of
PW's kinematic data, according to the results of a Kolmogorov-Smirnov
test (table \ref{tableAri}).
The radial extent of the gas in this model is larger than the other models 
(Figure 1b) and this leads to its producing more multiple hits 
along a given line of sight than any other model.

\begin{table}
\center
\caption{Five models of the gas distribution and their KS test results}
\label{tableAri}
\begin{tabular}{lclcccc}
Model & radial profile & normalization & $\Delta V$ & f$_{mm}$ & f$_{edg}$& f$_{tpk}$\\
\tableline
1   &  $e^{-R}$ &  $R_{gas}=R_{disk}$   & 0.001 & 0.008 & 0.001 & 0.001 \\
2   &  $e^{-R}$ &  $R_{gas}=7R_{disk}$  & 0.001 & 0.66 & 0.34 & 0.058 \\	
3   &  1/R	&  $N=0.59V_{c}/R$  &  0.001 & 0.004 & 0.001 & 0.001 \\
4   &  1/R      &  $R_{trunc}=R_{tidal} $  & 0.001 & 0.350 & 0.24 & 0.013  \\
5  & 1/R&  $log_{10}(N_{trunc})=19.3$ & 0.51 & 0.080 & 0.20 & 0.66 \\	
\end{tabular}
\end{table}
In fact the gas extends out beyond the tidal truncation radius 
(the limit of the fourth model), 
where it would no longer be gravitationally 
bound. 
However, our method of calculating the tidal radius is only 
approximate, and the large radial extent of the gas can perhaps
alternatively be interpreted as due to stripping or outflow.
If the gas is stripped away by ram-pressure stripping, or ejected from
the small proto-galaxies by supernovae, it may still cover an equivalent
amount of area, just not as a disk. We are investigating this more thoroughly.

Our main conclusion is that it is possible to produce the observed kinematic
properties of DLAS in SAMs based on CDM-type models 
by having lines of sight pass through multiple 
objects in the same halo; however, this requires that the cold gas in these
proto-galaxies be rather large in radial extent.

\begin{figure} 
\vskip .5pc
\begin{minipage}[b]{.46\linewidth}
\centering
\centerline{\epsfig{file=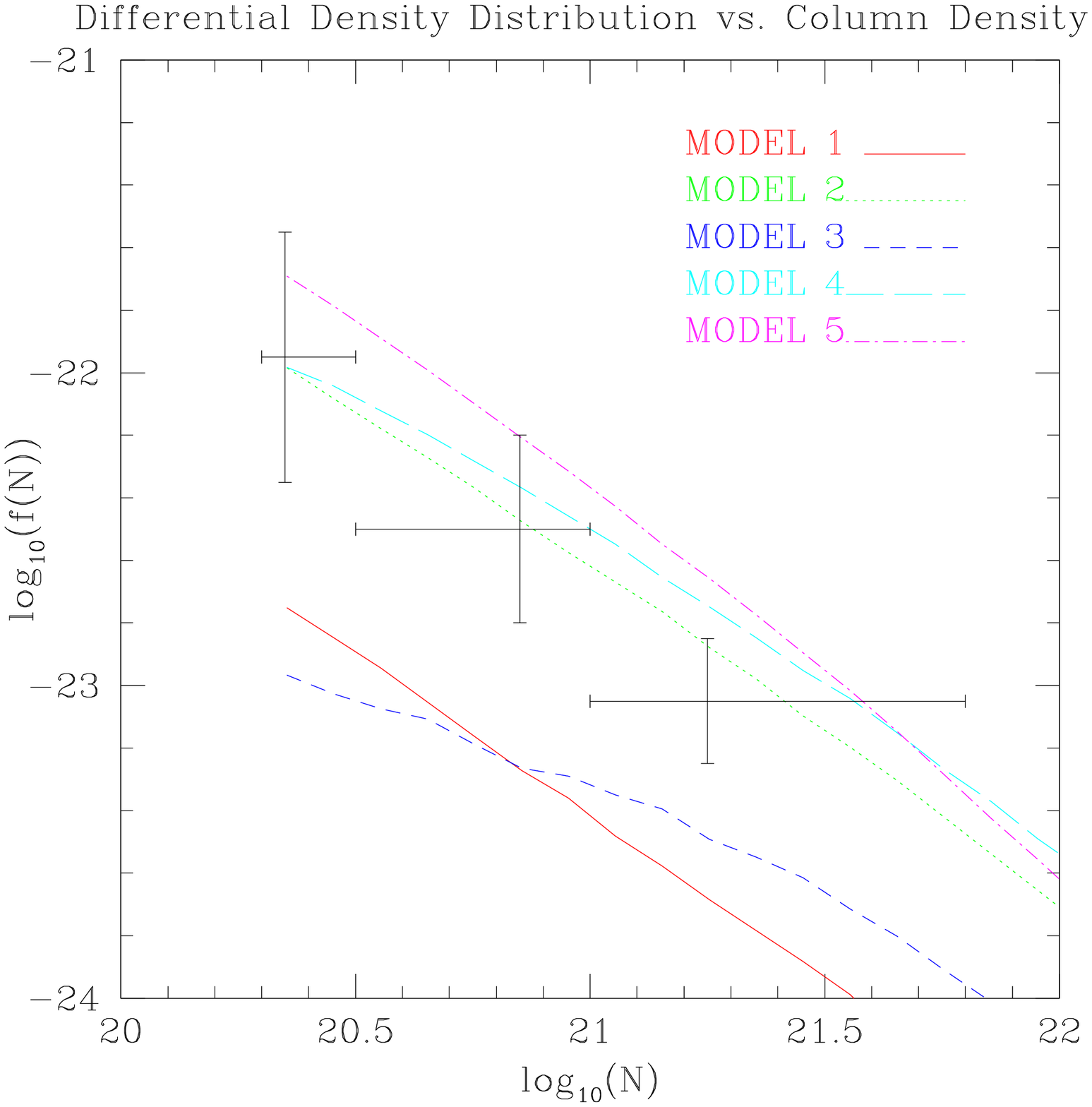,width=\linewidth}}
\end{minipage}\hfill
\begin{minipage}[b]{.46\linewidth}
\centering
\centerline{\epsfig{file=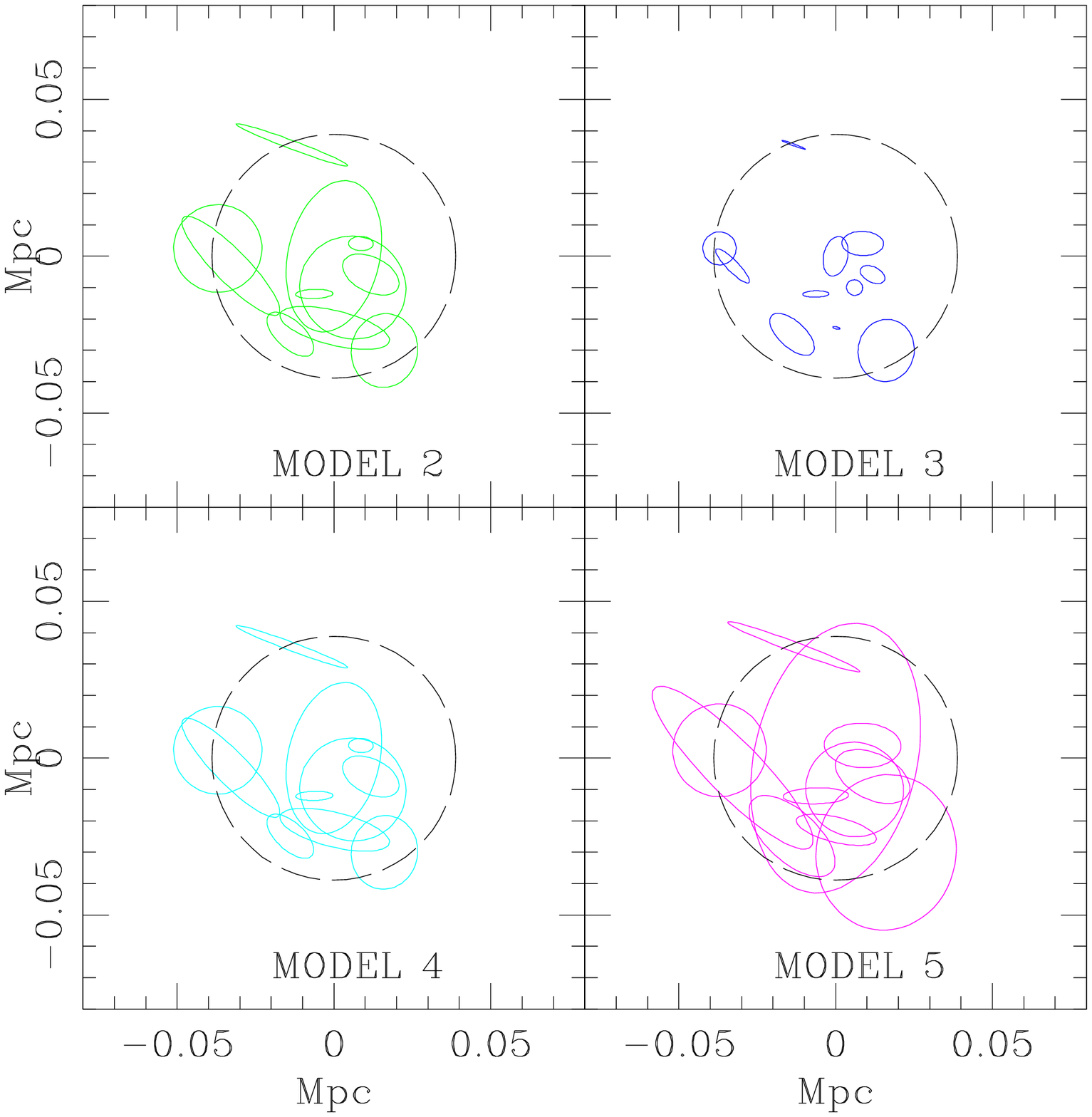,width=\linewidth}}
\end{minipage}\hfill
\vspace{10pt}
\caption{
a)The left panel shows the five models $f(N)$ distribution, 
compared to the data from
[3]. b)The right panel shows the extent of the gas disks in the models 2-5 
for a typical halo with circular velocity of 156kms$^{-1}$, 
the dashed line is the virial radius for this halo.
}\label{figure}
\end{figure}

\end{document}